\documentclass[
preprint,
superscriptaddress,
amsmath,
amssymb,
]{revtex4-1}

\usepackage{graphicx}
\usepackage{float}
\usepackage{dcolumn}
\usepackage{bm}
\usepackage{amsmath}
\usepackage{amssymb}
\usepackage{latexsym}
\usepackage{epsfig}
\usepackage{amsbsy}
\usepackage{array}
\usepackage{amssymb}
\usepackage{setspace}
\usepackage{bm}
\usepackage{indentfirst}
\usepackage{multirow}
\usepackage{color}

\begin{document}


\title{Relativistic, single-cycle tunable-infrared pulses generated by a tailored plasma density structure}


\author{Zan Nie}
\author{Chih-Hao Pai}
\email[]{chpai@tsinghua.edu.cn}
\author{Jianfei Hua}
\affiliation{Department of Engineering Physics, Tsinghua University, Beijing 100084, China}
\author{Chaojie Zhang}
\affiliation{University of California Los Angeles, Los Angeles, California 90095, USA}
\author{Yipeng Wu}
\author{Yang Wan}
\affiliation{Department of Engineering Physics, Tsinghua University, Beijing 100084, China}
\author{Fei Li}
\affiliation{Department of Engineering Physics, Tsinghua University, Beijing 100084, China}
\affiliation{University of California Los Angeles, Los Angeles, California 90095, USA}
\author{Zhi Cheng}
\author{Qianqian Su}
\author{Shuang Liu}
\author{Yue Ma}
\author{Xiaonan Ning}
\author{Yunxiao He}
\author{Wei Lu}
\email[]{weilu@tsinghua.edu.cn}
\affiliation{Department of Engineering Physics, Tsinghua University, Beijing 100084, China}
\author{Hsu-Hsin Chu}
\affiliation{Department of Physics, National Central University, Jhongli 32001, Taiwan}
\author{Jyhpyng Wang}
\affiliation{Department of Physics, National Central University, Jhongli 32001, Taiwan}
\affiliation{Institute of Atomic and Molecular Sciences, Academia Sinica, Taipei 10617, Taiwan}
\affiliation{Department of Physics, National Taiwan University, Taipei 10617, Taiwan}
\author{Warren B. Mori}
\affiliation{University of California Los Angeles, Los Angeles, California 90095, USA}
\author{Chan Joshi}
\affiliation{University of California Los Angeles, Los Angeles, California 90095, USA}


\date{\today}

\begin{abstract}
The availability of intense, ultrashort coherent radiation sources in the infrared region of the spectrum is enabling the generation of attosecond X-ray pulses via high harmonic generation, pump-probe experiments in the ``molecular fingerprint'' region and opening up the area of relativistic-infrared nonlinear optics of plasmas. These applications would benefit from multi-millijoule single-cycle pulses in the mid to long wavelength infrared (LW-IR) region. Here we present a new scheme capable of producing tunable relativistically intense, single-cycle infrared pulses from 5-14\,$\mu$m with a 1.7\% conversion efficiency based on a photon frequency downshifting scheme that uses a tailored plasma density structure. The carrier-envelope phase (CEP) of the LW-IR pulse is locked to that of the drive laser to within a few percent. Such a versatile tunable IR source may meet the demands of many cutting-edge applications in strong-field physics and greatly promote their development.
\end{abstract}


\maketitle


The past decade has seen a significant progress in the development of intense few-cycle mid-IR ($\lambda<5\,\mu$m) laser sources, which has opened new avenues in the research in strong-field laser-matter interaction\,\cite{wolter2015strong}. These mid-IR sources are an ideal tool for two-dimensional infrared spectroscopy\,\cite{calabrese2012ultrafast}, coherent soft X-ray high harmonic generation (HHG)\,\cite{popmintchev2012bright}, incoherent hard X-ray generation in laser-induced plasmas\,\cite{weisshaupt2014high}, and time-resolved imaging of molecular structures\,\cite{blaga2012imaging}. Furthermore, high-energy single-cycle mid-IR pulses can inherently isolate the electron dynamics in the strong-field interactions, which makes them very useful in investigating ultrafast phenomena in gases and solids. Some application examples are coherent control of lattice displacements through nonlinear photonics\,\cite{forst2011nonlinear}, the generation of isolated attosecond\,\cite{silva2015spatiotemporal}, or even zeptosecond X-ray pulses\,\cite{hernandez2013zeptosecond}, and sub-femtosecond control and metrology of bound-electron dynamics in atoms\,\cite{hassan2016optical}. In a number of these applications, such as incoherent hard X-ray generation\,\cite{weisshaupt2014high} and zeptosecond X-ray generation\,\cite{hernandez2013zeptosecond}, a long carrier wavelength ($\sim10\,\mu$m) is often preferred. However generation of high-energy few-cycle long-wavelength IR pulses is one of the current challenges to ultrafast laser technology. To date, the majority of high-intensity few-cycle mid-IR sources rely on parametric amplifiers (OPA) pumped by laser systems operating at $\sim1\,\mu$m\,\cite{andriukaitis201190, zhao2013generation}. However OPAs face significant challenges with an increase of wavelength beyond 5\,$\mu$m due to an increasingly unfavorable pump-to-idler photon energy ratio as well as a lack of nonlinear crystals with suitable optical and mechanical properties and/or transparency. Recently, there have been mid-IR optical parametric chirped pulse amplifier (OPCPA) demonstrations with sub-millijoule energy and wavelength reaching beyond 5\,$\mu$m in ZnGeP2 (ZGP) crystals\,\cite{von20175, sanchez20167}, pumped by high energy ~2\,$\mu$m picosecond Ho-doped lasers\,\cite{von2015picosecond, von2016ho, malevich2016broadband}, but the pulse duration is about 5-8 optical cycles. CO$_2$ lasers are important high-energy IR sources in the spectral range around $10\,\mu$m, but their wavelengths have limited tunability and their pulse durations are generally limited to a few picoseconds\,\cite{haberberger2010fifteen, polyanskiy2011picosecond}. Some other methods, such as  difference frequency generation (DFG)\,\cite{pupeza2015high}, adiabatic DFG\,\cite{krogen2017generation}, non-degenerate four-wave-mixing\,\cite{nomura2012phase, fuji2015generation, pigeon2016measurements} and optical rectification\,\cite{junginger2010single, silva2012multi, pigeon2014supercontinuum}, have attained microjoule level pulse energy, not sufficient for strong-field applications. There are also some post-processing methods, such as post compression\,\cite{mitrofanov2016subterawatt, shumakova2016multi} or synthesis\,\cite{liang2017high}, that are often more complex than direct generation methods. In other words, the generation of coherent, ultra-intense IR pulses is still an outstanding problem. In this paper we present an entirely different approach to solve this problem that utilizes asymmetric self-phase modulation (A-SPM) produced by a wake in a tailored (optimally shaped) plasma density structure such that the pump laser pulse continuously downshifts in frequency to give multi-millijoule energy, single-cycle long-wavelength IR (LW-IR) radiation. The advantage of this scheme for IR generation is that unlike in traditional nonlinear crystals, there is no damage limitation for the power scaling in plasmas.

\subsection*{Photon frequency down conversion in a nonlinear plasma wake}
\begin{figure*}[htbp]
\centering
  \includegraphics[width=0.9\textwidth]{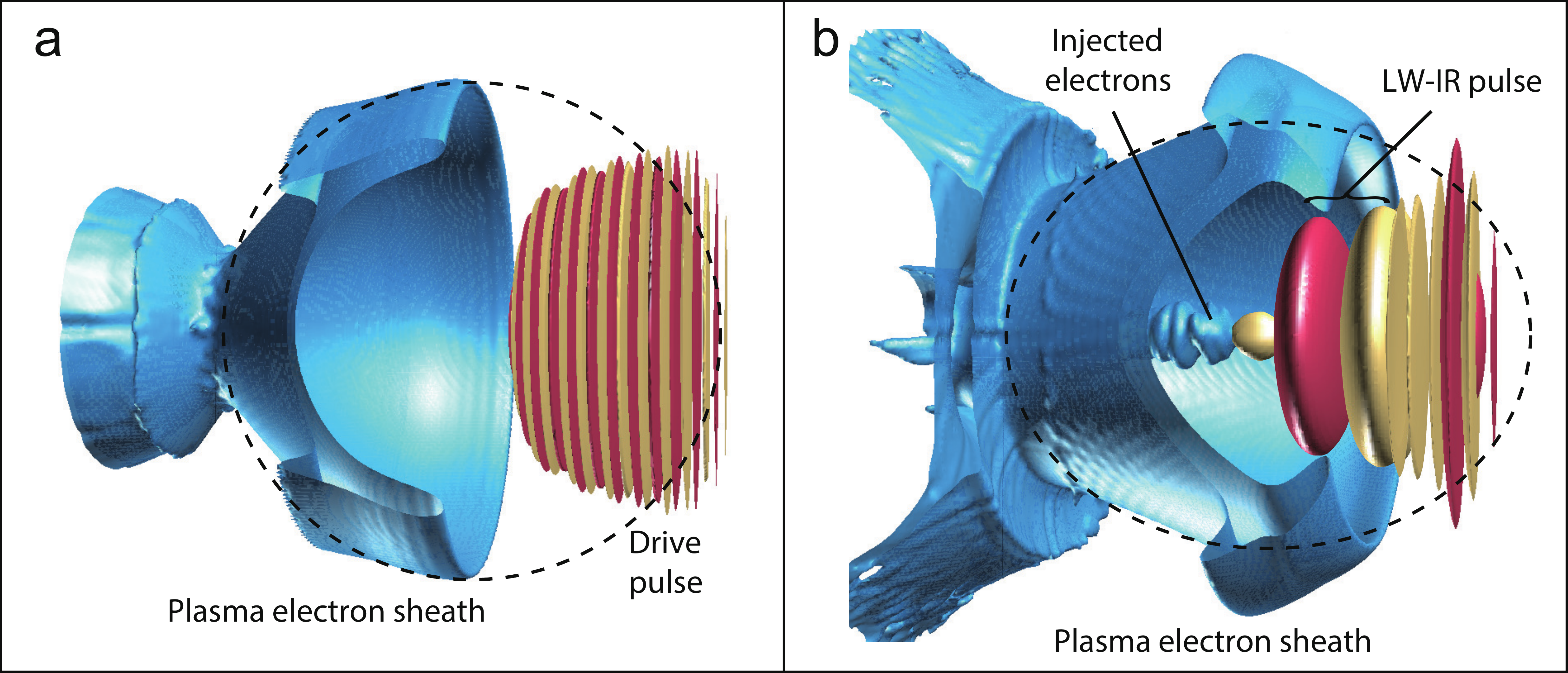}
   \caption{Illustration of photon frequency down conversion in a nonlinear plasma wake using a 3 dimensional particle-in-cell (PIC) simulation showing cut-away views of wakes. \textbf{a,} At the beginning of the structure where the laser pulse has formed the spherical wake (dotted circle) but has not undergone substantial frequency modification. \textbf{b,} Near the end of the tailored-density plasma structure where the laser pulse has been severely frequency downshifted and the LW-IR pulse is seen to reside in the elongated (dotted ellipse) wake cavity. Some plasma electrons are injected into the wake and accelerated by it.}
   \label{schematic_diagram}
\end{figure*}

It is well known that a portion of a short but intense laser pulse can be frequency downshifted rapidly as it excites a nonlinear wake (density disturbance) while propagating through an uniform underdense $(\omega_p<\omega_0)$ plasma\,\cite{gordon2003asymmetric, tsung2002generation}. Here $(\omega_p=4\pi n_e e^2/m_0)^{1/2}$ is the plasma frequency, $n_e$ is the ambient plasma density, and $\omega_0$ is the drive-laser¡¯s carrier frequency. If the normalized vector potential $a_0=eA/mc^2>2$, and the pulse duration $\tau<\sqrt2\pi /\omega_p$, the ponderomotive force (radiation pressure force) of the laser pulse eventually pushes out all the plasma electrons forward and outward, forming a 3D, nonlinear wake (a bubble-like region containing mostly plasma ions encapsulated in a sheath of plasma electrons) as shown in Fig.\,\ref{schematic_diagram}a. If $a_0>1$, the plasma electrons oscillate relativistically (close to c) in the laser electric field. According to 1D nonlinear theory, the index of refraction seen by the laser photons in the process varies as\,\cite{sprangle1990nonlinear0}: $\eta(t)\simeq 1-\frac{\omega_p^2}{2\omega_0^2}\frac{1}{1+\phi(t)}$ , where $\phi=|e|\Phi/m_0 c^2$ is the normalized scalar potential. $\eta$ varies because of the combined effects of the electron density variation and the relativistic variation of the electron mass\,\cite{sprangle1990nonlinear0, sprangle1990nonlinear}. It is the gradient of the refractive index (in the frame of the moving laser pulse: $\xi=z-ct$ ) that leads to a decrease or an increase in the frequency of the photons. Specifically, photons in the front of the wake where $\partial\eta/\partial\xi$ is negative are frequency downshifted, while photons overlapping the tail of the wake where $\partial\eta/\partial\xi$ is positive are frequency upshifted, and the part in the central, near-vacuum region comprising of plasma ions experiences scarcely any frequency change. In this respect the laser pulse undergoes frequency broadening due to SPM in the wake. This phenomenon has been well documented in the literature\,\cite{gordon2003asymmetric, wilks1989photon, esarey1990frequency, mori1997physics}. First effort to utilize this technique for generating ultrashort mid-IR pulses using a wake in a uniform plasma was reported in Ref.\,\cite{pai2010generation}, where about 1.5\% of the 810\,nm Ti:sapphire laser energy was converted into the wavelength range of 2--6\,$\mu$m.

In the tailored plasma density structure that we propose here, the laser pulse is initially compressed -- using a combination of SPM and group velocity dispersion (GVD) since $v_g\simeq c( 1-\frac{\omega_p^2}{2\omega_0^2}\frac{1}{1+\phi})$ -- and then rapidly further frequency downshifted in a higher density plasma region to generate the long-wavelength IR photons. The IR photons then slip back to the central region of the wake due to dispersion/slower group velocity. The central region of the wake is devoid of plasma electrons ($\eta \sim 1$) and thus serves as a perfect container for long-wavelength IR pulse, which can be seen in Fig.\,\ref{schematic_diagram}b. In this sense the present method can be thought of as an optimized case of extremely asymmetric self-phase modulation aided by GVD within a plasma wake.

\subsection*{Efficient single-cycle IR source using a tailored plasma density structure}
\begin{figure*}[htbp]
\centering
  \includegraphics[width=0.7\textwidth]{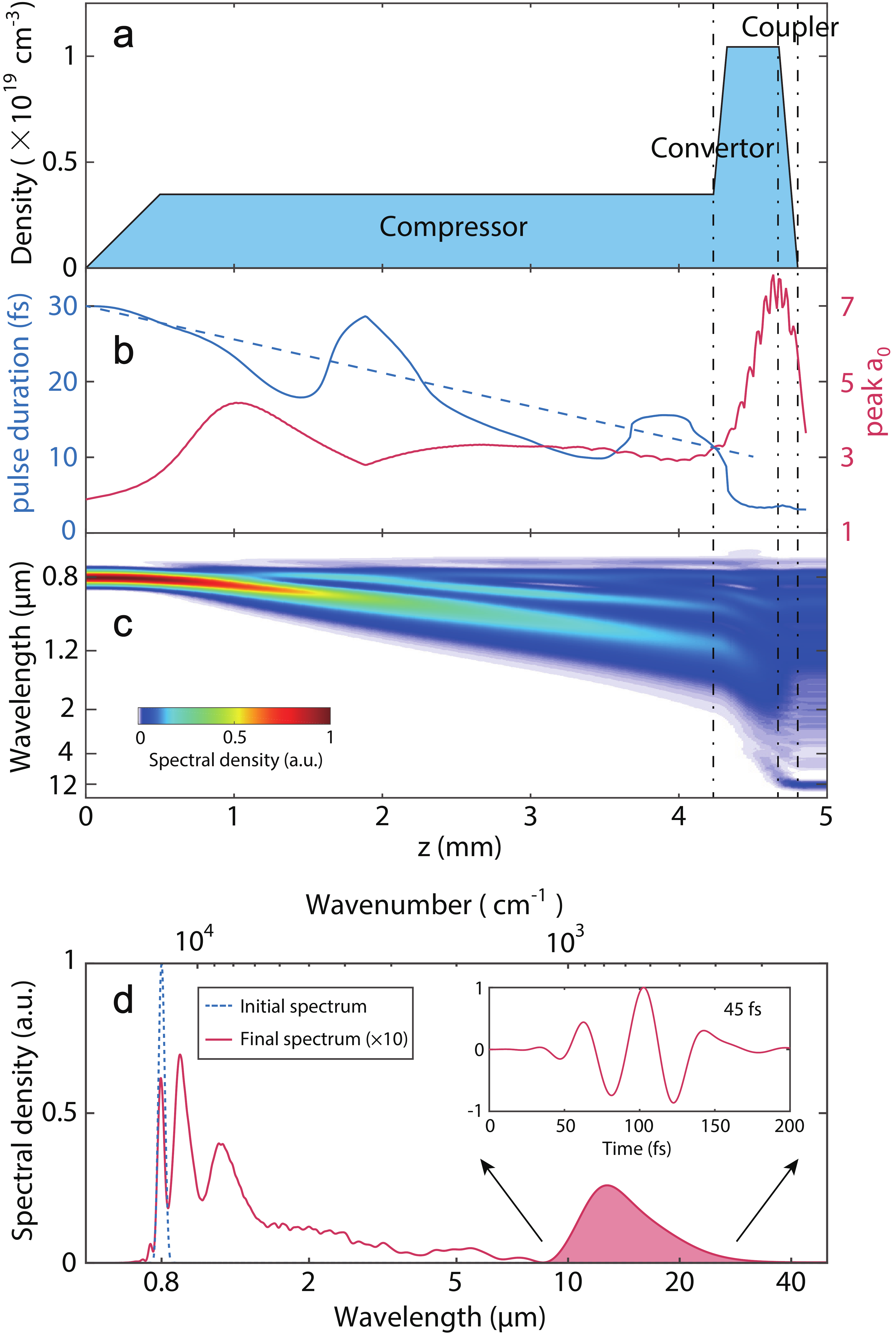}
   \caption{The structure and output of the LW-IR source. \textbf{a,} The density profile of the plasma structure. \textbf{b,} The pulse duration and peak a0 evolution with propagation distance in the plasma. The dotted blue line shows the general trend of the pulse duration variation in the compressor. \textbf{c,} The spectral evolution with the propagation distance in the plasma. \textbf{d,} The spectra of the laser pulse entering (blue dash) and exiting (red solid) the plasma structure and the IR pulse (shaded) that resides within the wake cavity shown in Fig.\,\ref{schematic_diagram}b. The inset shows the temporal variation of the electric field of the long wavelength portion (shaded) of the IR pulse. The plasma length is about 5mm or 5 Rayleigh lengths for the spot size we used. The wake encompasses much of the laser pulse and helps to guide it\,\cite{ralph2009self} over the entire length of the plasma structure.}
   \label{example}
\end{figure*}

As shown in Fig.\,\ref{schematic_diagram} for photon frequency down conversion to dominate the laser pulse evolution, it must be short enough (typically $<10$\,fs) so that the photons do work in pushing the plasma electrons outward. Such short pulses are often generated by pulse compression in hollow fibers\,\cite{guenot2017relativistic}, that can deliver $\sim$10\,mJ energy. To use such pulses to excite a nonlinear, fully blown out wake, high density (up to $10^{20}\,\text{cm}^{-3}$) plasmas have to be used. However in this case the LW-IR pulse cannot be generated since its wavelength is longer than the plasma wavelength and the pulse undergoes both photon up and down-shifting as explained earlier. Here we put forward a general solution that uses commonly available longer laser pulses (e.g., 30--50\,fs from a Ti-sapphire laser) with high enough energy to excite a fully blown out wake throughout the tailored plasma density structure. Our plasma structure consists of three different modules: the pulse compressor, the IR converter, and the output coupler as shown in Fig.\,\ref{example}a. Each module has a specific purpose that is accomplished by using a specific plasma density profile. First the pulse compressor shortens the 30--50\,fs long pulse that then undergoes rapid frequency downconversion in the converter section. The final coupler section transports the IR pulse out of the plasma efficiently while preserving its temporal quality.

We illustrate the physical processes occurring in each module of the plasma structure by showing the results obtained using the 3D-PIC simulation code OSIRIS\,\cite{fonseca2002osiris, fonseca2008one}. A laser pulse with an initial wavelength 800\,nm, pulse energy 1.2\,J, temporal FWHM of the $\text{sin}^2$ pulse of 30\,fs, and spot size 16\,$\mu$m is propagated through the plasma structure. The initial $a_0$ is about 2.1. The density profile of the plasma structure is shown in Fig.\,\ref{example}a. The FWHM pulse width and peak $a_0$ evolution is shown in Fig.\,\ref{example}b, the spectral evolution is shown in Fig.\,\ref{example}c and the initial spectrum and the final spectrum exiting the structure is shown in Fig.\,\ref{example}d with an inset indicating the temporal variation of the electric field of the long wavelength portion (shaded) of the IR pulse.
Here we point out that the plasma-accelerator community is already using similar tailored plasma density structures for the generation of high-brightness\,\cite{xu2017high} beams and for overcoming dephasing between the accelerating electrons and the plasma wake\,\cite{guillaume2015electron}. Structured density modules with the ratio of the density steps of factors of 2--3, peak densities on the order (1--2)$\times 10^{19}\,\text{cm}^{-3}$ and density ramps with scale lengths from 100--400\,$\mu$m have already been demonstrated. We now show that wake generation in such tailored density structures in addition can yield a tunable long-wavelength single-cycle IR source and explain the underlying physics of the process.

From Fig.\,\ref{example}b one can see that the laser pulse is compressed from 30\,fs to about 11\,fs in the compressor module and is further compressed to less than 4\,fs in the transition region between the compressor and converter modules. Because of the relatively low plasma density and the longer laser pulse length (compared to that in the converter module), the frequency spectrum of the pulse is only modestly broadened in the compressor module. Most of the frequency downconversion occurs in the higher density converter module. One can see this in Fig.\,\ref{example}c. The final coupler module simply transfers the pulse to the outside world with small attenuation and without any significant distortion due to GVD effects. The initial and final complete spectrum of the pulse is shown in Fig.\,\ref{example}d. The long wavelength IR portion appears as a distinct peak at the end of this spectrum (shaded region) and can be easily filtered. We find that this IR portion of the spectrum is nearly transform limited ($\Delta\nu\Delta\tau = 0.51$) and the oscillating field is only one period long as seen in the inset to Fig.\,\ref{example}d.

\begin{figure*}[htbp]
\centering
  \includegraphics[width=0.9\textwidth]{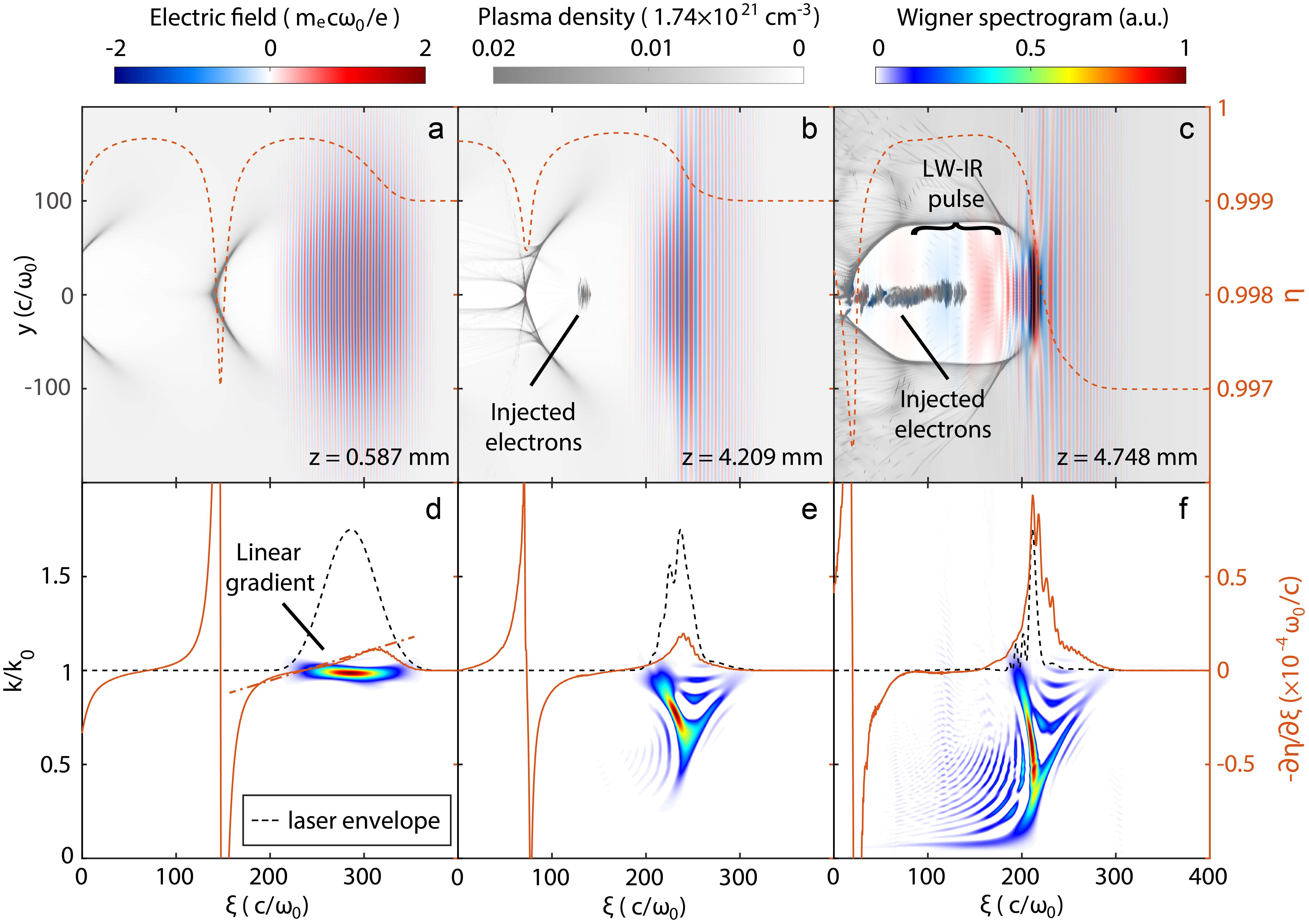}
   \caption{Photon frequency downconversion in a tailored plasma structure. The transverse electric field of the evolved laser pulse, the wake and on-axis refractive index (orange dash) are shown (\textbf{a}) at the beginning of the uniform density part of the pulse compressor; (\textbf{b}) at the end of the pulse compressor, and (\textbf{c}) at the end of the IR converter. The Wigner spectrograms of the on-axis transverse electric field, the gradient of refractive index (orange line), and laser envelope (black dash line) are shown in (\textbf{d}), (\textbf{e}) and (\textbf{f}) at the same position as in (\textbf{a}), (\textbf{b}) and (\textbf{c}) respectively. In (\textbf{b}) and (\textbf{c}) the injection of some plasma electrons is also seen.}
   \label{example_wigner}
\end{figure*}

\emph{\textbf{1) Compressor module:}} Now we discuss in detail the underlying physics in each module of the plasma structure. In the pulse compressor module, the plasma density is relatively low (3.5$\times 10^{18}\,\text{cm}^{-3}$). The laser pulse creates a nonlinear wake as it enters the plasma. The varying refractive index gradient due to the wake formation causes the instantaneous frequency to vary locally along laser pulse. The whole laser pulse undergoes frequency downshifting and develops a nearly linear chirp -- because of an almost linear refractive index gradient at the beginning of the compressor (dashed-dotted line in Fig.\,\ref{example_wigner}d) over most of the pulse -- which is optimal for self-compression by GVD. However, with the process of self-compression, the nearly linear refractive index gradient gradually gets nonlinear (Fig.\,\ref{example_wigner}e). This leads to a higher order chirp, and produces the variation of pulse duration (Fig.\,\ref{example}b). Overall, the negative GVD of the plasma results in self-compression of the laser pulse from 30\,fs to 11\,fs (Fig.\,\ref{example_wigner}b) so that it has a duration needed for IR generation in the converter module (Fig.\,\ref{example_wigner}c).

\emph{\textbf{2) Converter module:}} The 1D quasi-static nonlinear theory\,\cite{sprangle1990nonlinear0} can be used to design the laser and plasma parameters that are optimum for the converter module (see Methods). In this module, the plasma density is three times higher than that in the pulse compressor. The converter length is only 450\,$\mu$m including a 100\,$\mu$m plasma density up-ramp that serves as the transition region. The increase of plasma density leads to self-focusing and further self-compression of the drive pulse. This together with frequency downshifting of the photons leads to a rapid increase of $a_0$ of the pulse (from 2.1 to 7.8 as shown in Fig.\,\ref{example}b since $a_0\propto E/\omega$). Throughout this module the whole drive pulse resides in the very front of the wake where it experiences an extremely rapid change of the index of refraction as shown in Fig.\,\ref{example_wigner}c, f. Interestingly, the longest-wavelength IR components slip rapidly backwards towards the center of the wake due to their much slower group velocity. In the almost vacuum-like region (devoid of plasma electrons) in this part of the wake, the IR pulse undergoes little further photon frequency change or dispersion. Meanwhile, the wake wavelength is elongated ($\lambda_{\text{nl}}=\sqrt{a_0}\cdot 2\pi c/\omega_p$) due to an increase of $a_0$, which increases the volume of the ion cavity region so that it can accommodate the long-wavelength IR pulse.

The scenario described above is confirmed in Fig.\,\ref{example_wigner}d--f which show the Wigner transforms (instantaneous wavenumber (frequency) as a function of time) of the on axis electric field superimposed on the variation of the (negative) gradient of the refractive index  ($-\partial\eta/\partial\xi$) at these same positions. As expected when $\partial\eta/\partial\xi$ is negative (positive) the frequency is downshifted (upshifted) from the definition of instantaneous frequency as the rate of change of phase and thus the refractive index. At the beginning of the uniform density part of the compressor, the initially sin2 pulse occupies mostly the first half of the wake (Fig.\,\ref{example_wigner}a) and therefore undergoes SPM that leads to gentle frequency downshift at a different rate in different parts of the pulse (Fig.\,\ref{example_wigner}d). By the end of the compressor section the pulse is compressed (Fig.\,\ref{example_wigner}b) and the peak of the laser pulse is now located where the refractive index gradient is the largest (Fig.\,\ref{example_wigner}e). At the end of the IR converter module, the pulse is rapidly downshifted with the broad infrared spectrum forming a very narrow pulse at the front followed by the long wavelength IR pulse that slips back and resides in the wake cavity (Fig.\,\ref{example_wigner}c, f). The total conversion efficiency of the LW-IR radiation in the range of 8--30\,$\mu$m is about 1.7\%. However all these frequency components are phase locked to give a single cycle pulse with a carrier wavelength of 12\,$\mu$m. In other words, about 20\,mJ (0.44\,TW) of energy is contained in an $\sim$1 wavelength cube volume leading to a relativistically intense pulse with an $a_0$ of $>$ 7.

\begin{figure*}[htbp]
\centering
  \includegraphics[width=0.5\textwidth]{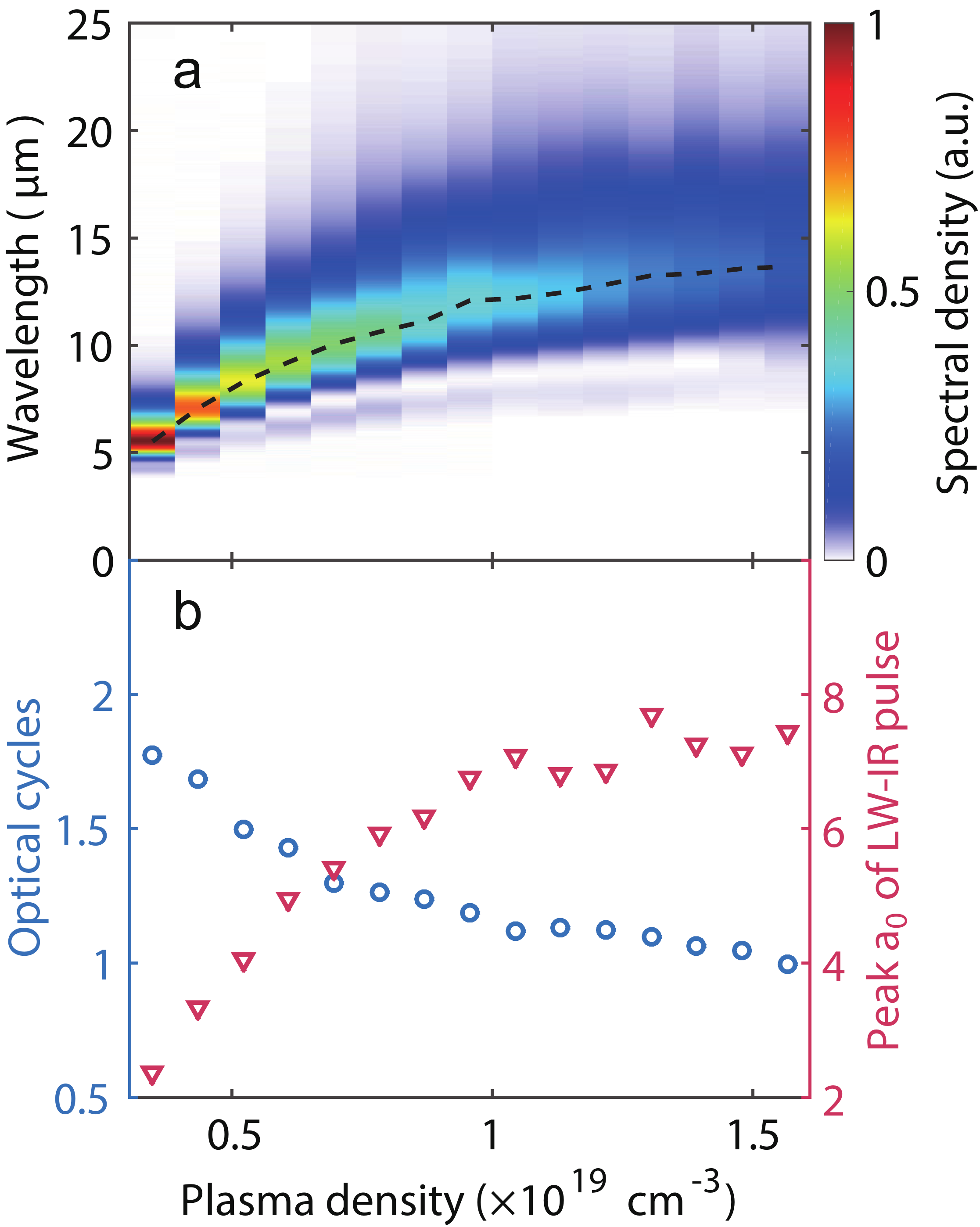}
   \caption{Effect of variation of plasma density on output parameters of the LW-IR pulse in the converter module. Dependence of (\textbf{a}) the IR spectrum and (\textbf{b}) optical cycles and peak $a_0$ on plasma density. The dashed line in (\textbf{a}) tracks the carrier wavelength of the LW-IR pulse.}
   \label{density_spectrum}
\end{figure*}

We have scanned the plasma and the laser pulse parameters to investigate how robust the results shown in Fig.\,\ref{example} and \ref{example_wigner} are. For this we use the quasi-3D OSIRIS code\,\cite{lifschitz2009particle, davidson2015implementation} to save simulation time. We find that the plasma density is a decisive factor for determining the central wavelength of the IR pulse. By varying plasma density and plasma length in the IR converter module, the central wavelength of the IR pulse can be tuned from 5\,$\mu$m to 14\,$\mu$m (Fig.\,\ref{density_spectrum}a). We find that for plasma densities higher than 1$\times 10^{19}\,\text{cm}^{-3}$, it is possible to obtain near single-cycle LW-IR radiation of variable wavelength (Fig.\,\ref{density_spectrum}b). The density however cannot be made arbitrarily high because the pulse must reside in the portion of the wake where $\partial\eta/\partial\xi$ is negative. Furthermore, if the plasma density in the converter module is too high the drive pulse may not be able to sustain a large enough cavity to contain the generated IR pulse, resulting in significant attenuation. Besides, the scale length of the downramp must be shorter the higher the plasma density. This proves to be a practical limitation.

\begin{figure*}[htbp]
\centering
  \includegraphics[width=0.9\textwidth]{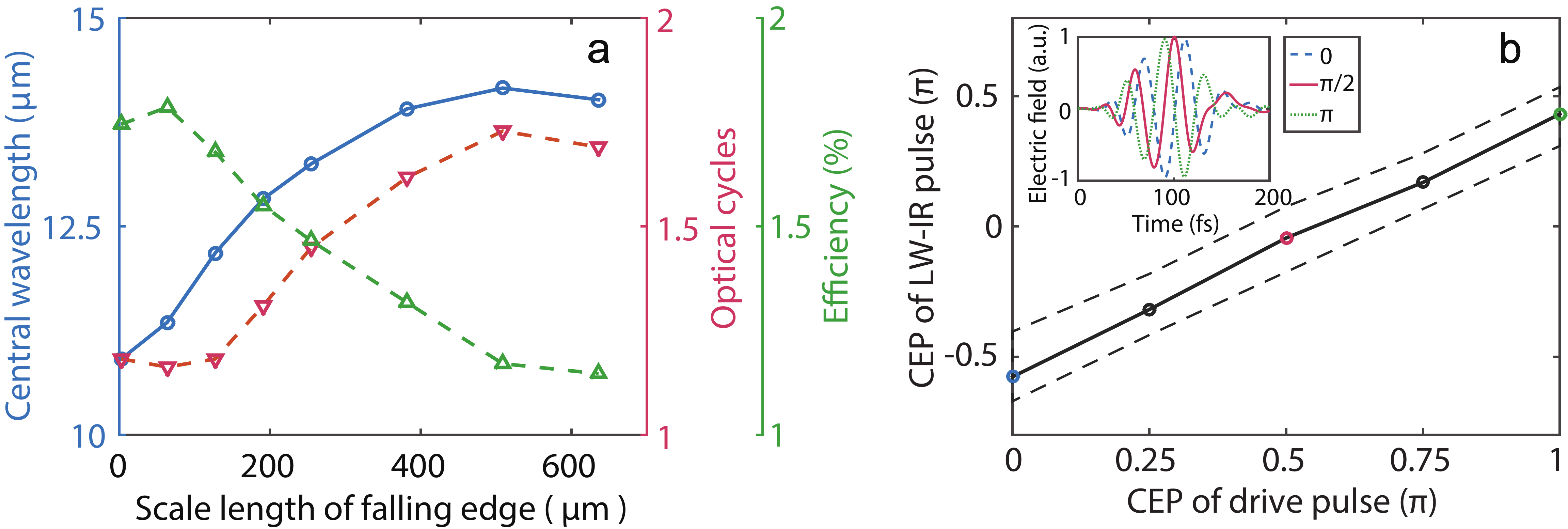}
   \caption{\textbf{a,} Dependence of central wavelength, optical cycles, and IR conversion efficiency (8-30\,$\mu$m) on scale length of falling edge in the output coupler. \textbf{b,} The relationship between CEP of the LW-IR pulse and that of the initial drive laser pulse for the case shown in Fig.\,\ref{example}. The upper (lower) bound of the shaded region is the result of increasing (decreasing) the integral of plasma density and length by 3.3$\times 10^{19}\,\text{cm}^{-3}\cdot \mu$m. The inset shows the electric field of filtered LW-IR pulse for different initial CEP of the drive laser (blue dash -- 0; red line -- $\pi$/4; green dot -- $\pi$/2).}
   \label{downramp_and_CEP}
\end{figure*}

\emph{\textbf{3) Coupler module:}} After the LW-IR pulse is generated in the converter module, it has to be extracted from the plasma without either dispersive spreading or losing energy due to absorption. The closer the IR frequency to the initial plasma frequency the worse is the attenuation. The coupler solves this problem by expanding the wake cavity size in the density downramp. This is so because nonlinear wake wavelength (cavity size) is given by $\lambda_{\text{nl}}\propto \sqrt{a_0}/\omega_p$ and in the down-ramp $a_0$ (Fig.\,\ref{example}b) falls slower than $\omega_p$. The programmable parameter here is the scale length $L=(\frac{1}{n_e}\frac{dn_e}{dx})^{-1}$ of the falling plasma density $n_e(x)$. The scale length of falling edge must be short enough so that the bubble expands quickly and the center of the wake cavity retreats very quickly relative to the IR pulse, so that the IR pulse leaves the plasma with neither energy loss nor further frequency change. Otherwise, the center of the wake still retreats relative to the IR pulse but with a slower velocity. In this case, the IR pulse can initially experience further frequency downconversion since it resides in the front of the wake again due to the retreat of the wake center. However, the increase of the downramp length will lead to longer IR pulse duration due to GVD. In Fig.\,\ref{downramp_and_CEP}a we show how the central wavelength, the optical cycles, and the conversion efficiency of the LW-IR pulse are affected by the scale-length of the downramp of the coupler module for the case considered here. With the increase of the scale length of the falling edge, the central wavelength increases, the optical cycles also increase, and the conversion efficiency drops. Taking all these factors into account, the optimal scale length of falling edge for the example shown in Fig.\,\ref{example} is about 130\,$\mu$m.

\textbf{Carrier-envelope phase locking:} We find that the carrier envelope phase (CEP) of the IR pulse generated in this scheme is locked to the initial CEP of the drive laser pulse. This property is especially desirable to multi-shot pump probe applications where the single cycle IR pulse is used as the pump and a small fraction of the initial laser pulse is used as a probe and the delay between the two is varied on micrometer scale.

When a transform limited laser pulse propagates in plasma, the change of CEP is inevitable since the laser phase (carrier) velocity $v_p$  is faster than the speed of light while its group (envelope) velocity $v_g$ is slower than the speed of light. The change in the phase of the electric field between the LW-IR pulse and the initial drive pulse measured at the peak intensity point of the two pulses is given by:
\begin{align}
\Phi_{\text{CEP}}-\Phi_0=\int_{0}^{l/c} \frac{2\pi (v_p-v_g )dt}{\lambda_{\text{IR}}(t)} \simeq \int_{0}^{l/c} \frac{n_e(t)}{n_c} \frac{\omega_0^2}{\omega_{\text{IR}}(t)} \frac{dt}{1+\phi(t)} \,, \label{eq:CEP}
\end{align}
where $\Phi_0$ is the initial CEP of the drive pulse, $\omega_0$ is the initial drive pulse frequency, $n_e(t)$ is the ambient plasma density, $\omega_{\text{IR}}(t)$ is the IR pulse frequency, and $l$ is the total length of the plasma. Once the plasma density profile is fixed, the variation of CEP caused by propagation is fixed. Then there is a linear mapping relationship between the CEP of generated IR pulse and that of the initial drive laser pulse. We use the same parameters as those used in Fig.\,\ref{example} to illustrate this mapping. In Fig.\,\ref{downramp_and_CEP}b we plot the CEP of the $\lambda=12\,\mu$m IR pulse, as the CEP of the drive pulse is varied from 0 to $\pi$ (for $\lambda=0.8\,\mu$m) radians. The upper (lower) dashed black line in Fig.\,\ref{downramp_and_CEP}b shows the resulting CEP of the IR pulse for the same drive pulse but with the plasma density - length product increased (decreased) by $3.3\times 10^{19}\,\text{cm}^{-3}\cdot \mu$m. We can see that the two track one another rather well in all three cases.

Based on Eq.\,1, we can roughly estimate the CEP evolution at different sections. For the pulse compressor section, $\omega_{\text{IR}}(t)$ changes little due to gentle photon frequency downshifting (Fig.\,\ref{example}c), and $\phi(t)$ also changes little due to little change of $a_0$ in this process (Fig.\,\ref{example}b). Therefore, the variation of CEP is roughly dependent on the integral of the plasma density and length. For the IR converter section, $\omega_{\text{IR}}(t)$ changes rapidly due to strong photon frequency downshifting (Fig.\,\ref{example}c), and $\phi(t)$ also changes rapidly due to rapid change of $a_0$ in this process (Fig.\,\ref{example}b). Interestingly, both these changes approximately compensate each other so that the variation of CEP is still roughly dependent on the integral of the plasma density and length. For solid state CEP-stable lasers, the CEP jitter is often less than $ 0.1\pi$ radians. As shown in Fig.\,\ref{downramp_and_CEP}b, to achieve the same degree of control, the jitter of the plasma density - length product should be controlled less than $3.3\times 10^{19}\,\text{cm}^{-3}\cdot \mu$m, since these two parameters are linearly correlated. Besides, while keeping the integral of the plasma density and length unchanged, we have changed the scale length of the upramp between the compressor and the converter in the range of 100--250\,$\mu$m, the CEP only changes about 0.02$\pi$, which supports the physical explanation given above.

\subsection*{Conclusion}
In summary, we have developed and analyzed a scheme that generates intense long-wavelength single-cycle IR pulses by frequency downshifting in a tailored density plasma structure and demonstrated it using 3D PIC simulations. Our study shows this source is able to provide single cycle, high-peak-power tunable IR pulses in a spectral region $5< \lambda(\mu \text{m})<14$ where no such source currently exists.

\section*{Methods}
\subsection*{Theoretical estimate of the optimal pulse duration and plasma length of the IR converter}
We use 1D quasi-static nonlinear theory to estimate the optimal pulse duration and the length of the IR converter. The evolution of the vector potential of the laser pulse and the scalar potential of the wake are given by the following coupled equations\,\cite{sprangle1990nonlinear0}:

\begin{align}
&\left[-\frac{2}{c}\frac{\partial}{\partial\xi}-\frac{1}{c^2}\frac{\partial}{\partial\tau}\right]
\frac{\partial \boldsymbol{a}}{\partial \tau}=k^2_p\frac{\boldsymbol{a}}{1+\phi}\,,\\
&\frac{\partial^2 \phi}{\partial \xi^2}=\frac{k^2_p}{2}\left[\frac{1+a^2}{(1+\phi)^2}-1 \right] \,, \label{eq:phi equation}
\end{align}
where $\xi=ct-z,\tau=t$ are the speed of light coordinates, $\phi=|e|\Phi/m_0 c^2$ is the normalized scalar potential. When the normalized scalar potential $|\phi|\ll1$, it can be expressed as\,\cite{sprangle1990nonlinear}:
\begin{align}
\phi \simeq \frac{k_p}{2}\int_{\xi}^{0}a_L^2(\xi')sin[k_p(\xi'-\xi)]d\xi'\,,
\end{align}
where $a_L$ is the envelope amplitude of the normalized vector potential $\boldsymbol{a}$, and $k_p=\omega_p/c$, $\omega_p=(4\pi |e|^2n_e/m_0)^{1/2}$ is the ambient plasma frequency, $n_e$ is the ambient plasma density. If the pulse envelope is given by $a_L=a_0sin^2(\pi\xi/L)$ for $-L\leq\xi\leq0$, we can get
\begin{align}
\phi \simeq \left(\frac{p}{16}\right)^2 \left\{ 12\theta^2-\left[1-cos(2\theta)\right] \cdotp \left[7-cos(2\theta)\right] \right\} \,,\label{eq:phi2}
\end{align}
where $p=k_p a_0 L/\pi$, $\theta=\pi\xi/L$. The local wavenumber of the laser pulse is modified during the interaction with the plasma due to the gradient of the index of refraction, which can be expressed by $\phi$ in 1D theory. Then the photon deceleration rate is expressed as\,\cite{mori1997physics}:
\begin{align}
r\equiv\frac{1}{k}\frac{\partial k}{\partial (c\tau)}\simeq-\frac{1}{2}\frac{k_p^2}{k^2}\frac{1}{(1+\phi)^2}\frac{\partial\phi}{\partial\xi}= -\frac{\pi}{2L}\frac{k_p^2}{k^2}\frac{1}{(1+\phi)^2}\frac{\partial\phi}{\partial\theta} \,. \label{eq:k}
\end{align}

\begin{figure*}[htbp]
\centering
  \includegraphics[width=0.9\textwidth]{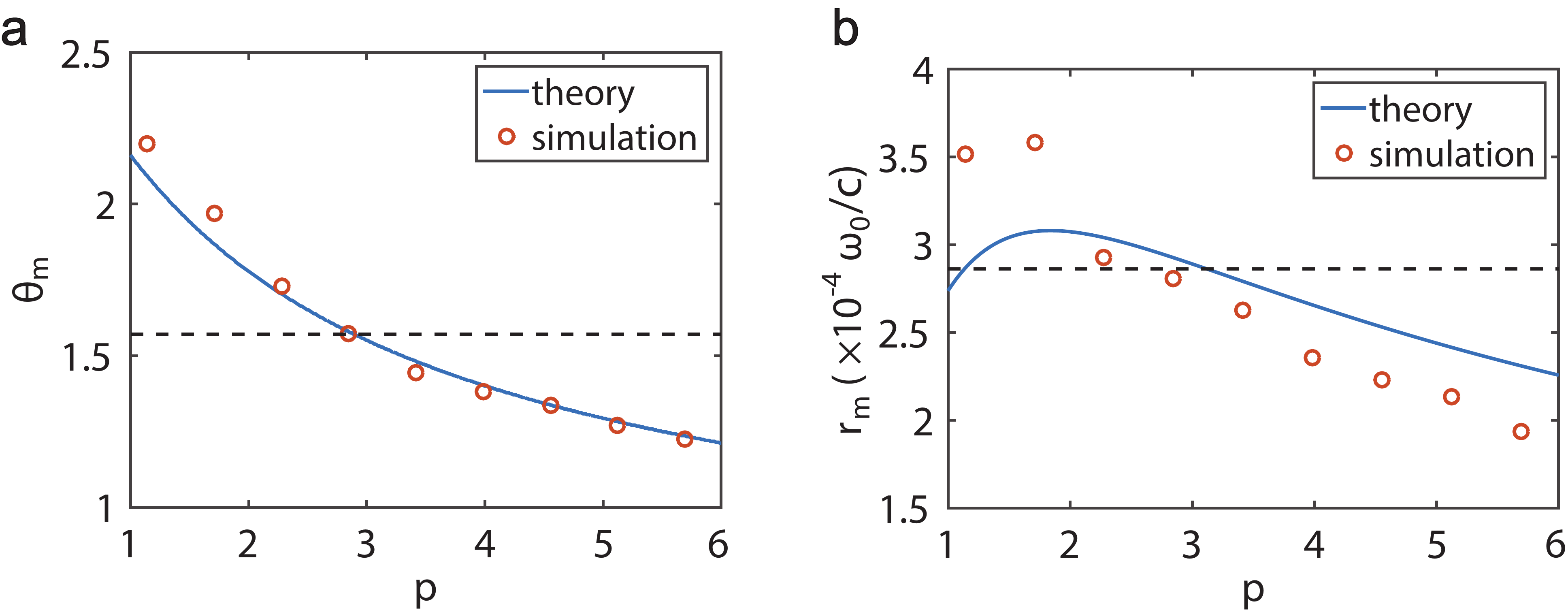}
   \caption{Comparison of theory results and simulation results. \textbf{a,} Dependence of position of the maximum photon deceleration rate on parameter $p$. \textbf{b,} Dependence of maximum photon deceleration rate on parameter $p$.}
   \label{thetam}
\end{figure*}

Inserting Eq.\,\ref{eq:phi2} into Eq.\,\ref{eq:k}, we can get the photon frequency downshifting rate. From Eq.\,\ref{eq:phi2}, we know that for a given parameter $p$, the profile of the photon frequency downshifting rate is determined and accordingly the position of the maximum photon frequency downshifting rate $\theta_m$ (extreme point) is determined. Then there is a certain relationship between them, that is $\theta_m=f(p)$, which is shown in Fig.\,\ref{thetam}a.

When $\theta_m=\pi/2$,  (dashed line in Fig.\,\ref{thetam}a) the position of the maximum photon frequency downshifting rate overlaps with the peak of the laser pulse. Numerically, we find that the optimal $p=2.88$ (Fig.\,\ref{thetam}a), which agrees very well with simulation results. Then for given plasma density and normalized vector potential of drive pulse, the optimal drive pulse length can be obtained:
\begin{align}
L=\frac{1.44\lambda_p}{a_0} \,.
\end{align}
where $\lambda_p$ is the plasma wavelength. Here, $a_0\gg 1$ should be satisfied to make sure that the premise of the derivation ($L \ll \lambda_p$) is still valid. To get a comparable parameter with experiments, the pulse duration (FWHM of intensity) is given by:
\begin{align}
c\tau=0.36L=\frac{0.52\lambda_p}{a_0} \,.
\end{align}

Inserting $\theta_m=\pi/2$ into Eq.\,\ref{eq:phi2} and Eq.\,\ref{eq:k}, the maximum photon deceleration rate $r_m$ is obtained (assuming the optimal pulse duration):
\begin{align}
r_m \simeq-0.1\frac{k_p^{3}}{k^2}a_0\,.\label{eq:rate}
\end{align}
If $a_0=const$, the laser wavenumber can be written as:
\begin{align}
k^2 \simeq k_0^2-0.2k_p^3a_0l\,,
\end{align}
where $k_0$ is the initial wavenumber, $l$ is the propagation distance in a plasma. In the case of large frequency downshifting, $k\ll k_0$, then we can get the optimal plasma length $l_m$ to obtain LW-IR pulse:
\begin{align}
l_m\simeq \frac{5k_0^2}{k_p^3}a_0^{-1} \,.\label{eq:plasma_length}
\end{align}
In practice the whole pulse cannot undergo photon frequency downshifitng with the maximum rate, and therefore the plasma length should be increased by up to a factor of 2. In addition, $a_0$ changes a lot due to strong self-focusing and self-compression in high density plasmas. Therefore, $a_0$ is an average value over time in Eq.\,\ref{eq:plasma_length}.

\subsection*{Particle-in-cell simulation}
The 3D PIC simulations were carried out using the code OSIRIS\,\cite{fonseca2002osiris, fonseca2008one} in Cartesian coordinates with a window moving at the speed of light. The z axis was defined to be the drive laser propagating direction. The simulation window had a dimension of $102\times 102\times 61\mu$m with $600\times 600\times 2400$ cells in the x, y, and z directions, respectively. This corresponded to cell sizes of $\Delta x=\Delta y=1.33k_0^{-1}$ and $\Delta z=0.2k_0^{-1}$ (where $k_0=2\pi\lambda_0^{-1}$ is the laser wavevector and $\lambda_0=800\,$nm. The number of macro-electrons per cell was 2.

The parameter scan was carried out using the quasi-3D OSIRIS\,\cite{lifschitz2009particle, davidson2015implementation} to save simulation time. Simulations were performed using a mesh with $\Delta r=k_0^{-1}$ and $\Delta z=0.2k_0^{-1}$ and the first two Fourier modes. The number of macro-electrons per cell was 4.

\bibliography{refs}

\begin{thebibliography}{10}

\bibitem{wolter2015strong}
Benjamin Wolter, Michael~G Pullen, Matthias Baudisch, Michele Sclafani,
  Micha{\"e}l Hemmer, Arne Senftleben, Claus~Dieter Schr{\"o}ter, Joachim
  Ullrich, Robert Moshammer, and Jens Biegert.
\newblock Strong-field physics with mid-ir fields.
\newblock {\em Phys. Rev. X}, 5(2):021034, 2015.

\bibitem{calabrese2012ultrafast}
Carmella Calabrese, Ashley~M Stingel, Lei Shen, and Poul~B Petersen.
\newblock Ultrafast continuum mid-infrared spectroscopy: probing the entire
  vibrational spectrum in a single laser shot with femtosecond time resolution.
\newblock {\em Opt. Lett.}, 37(12):2265--2267, 2012.

\bibitem{popmintchev2012bright}
Tenio Popmintchev, Ming-Chang Chen, Dimitar Popmintchev, Paul Arpin, Susannah
  Brown, Skirmantas Ali{\v{s}}auskas, Giedrius Andriukaitis, Tadas
  Bal{\v{c}}iunas, Oliver~D M{\"u}cke, Audrius Pugzlys, et~al.
\newblock Bright coherent ultrahigh harmonics in the kev x-ray regime from
  mid-infrared femtosecond lasers.
\newblock {\em Science}, 336(6086):1287--1291, 2012.

\bibitem{weisshaupt2014high}
Jannick Weisshaupt, Vincent Juv{\'e}, Marcel Holtz, ShinAn Ku, Michael Woerner,
  Thomas Elsaesser, Skirmantas Ali{\v{s}}auskas, Audrius Pug{\v{z}}lys, and
  Andrius Baltu{\v{s}}ka.
\newblock High-brightness table-top hard x-ray source driven by
  sub-100-femtosecond mid-infrared pulses.
\newblock {\em Nat. Photonics}, 8(12):927--930, 2014.

\bibitem{blaga2012imaging}
Cosmin~I Blaga, Junliang Xu, Anthony~D DiChiara, Emily Sistrunk, Kaikai Zhang,
  Pierre Agostini, Terry~A Miller, Louis~F DiMauro, and CD~Lin.
\newblock Imaging ultrafast molecular dynamics with laser-induced electron
  diffraction.
\newblock {\em Nature}, 483(7388):194--197, 2012.

\bibitem{forst2011nonlinear}
Michael F{\"o}rst, Cristian Manzoni, Stefan Kaiser, Yasuhide Tomioka, Yoshinori
  Tokura, Roberto Merlin, and Andrea Cavalleri.
\newblock Nonlinear phononics as an ultrafast route to lattice control.
\newblock {\em Nat. Phys.}, 7(11):854--856, 2011.

\bibitem{silva2015spatiotemporal}
Francisco Silva, Stephan~M Teichmann, Seth~L Cousin, Michael Hemmer, and Jens
  Biegert.
\newblock Spatiotemporal isolation of attosecond soft x-ray pulses in the water
  window.
\newblock {\em Nat. Commun.}, 6, 2015.

\bibitem{hernandez2013zeptosecond}
C~Hern{\'a}ndez-Garc{\'\i}a, JA~P{\'e}rez-Hern{\'a}ndez, T~Popmintchev,
  MM~Murnane, HC~Kapteyn, A~Jaron-Becker, A~Becker, and L~Plaja.
\newblock Zeptosecond high harmonic kev x-ray waveforms driven by midinfrared
  laser pulses.
\newblock {\em Phys. Rev. Lett.}, 111(3):033002, 2013.

\bibitem{hassan2016optical}
M~Th Hassan, Tran~Trung Luu, Antoine Moulet, O~Raskazovskaya, P~Zhokhov, Manish
  Garg, Nicholas Karpowicz, AM~Zheltikov, V~Pervak, Ferenc Krausz, et~al.
\newblock Optical attosecond pulses and tracking the nonlinear response of
  bound electrons.
\newblock {\em Nature}, 530(7588):66, 2016.

\bibitem{andriukaitis201190}
Giedrius Andriukaitis, Tadas Bal{\v{c}}i{\=u}nas, Skirmantas Ali{\v{s}}auskas,
  Audrius Pug{\v{z}}lys, Andrius Baltu{\v{s}}ka, Tenio Popmintchev, Ming-Chang
  Chen, Margaret~M Murnane, and Henry~C Kapteyn.
\newblock 90 gw peak power few-cycle mid-infrared pulses from an optical
  parametric amplifier.
\newblock {\em Opt. Lett.}, 36(15):2755--2757, 2011.

\bibitem{zhao2013generation}
Kun Zhao, Haizhe Zhong, Peng Yuan, Guoqiang Xie, Jing Wang, Jingui Ma, and
  Liejia Qian.
\newblock Generation of 120 gw mid-infrared pulses from a widely tunable
  noncollinear optical parametric amplifier.
\newblock {\em Opt. Lett.}, 38(13):2159--2161, 2013.

\bibitem{von20175}
Lorenz von Grafenstein, Martin Bock, Dennis Ueberschaer, Kevin Zawilski, Peter
  Schunemann, Uwe Griebner, and Thomas Elsaesser.
\newblock 5 $\mu$m few-cycle pulses with multi-gigawatt peak power at a 1 khz
  repetition rate.
\newblock {\em Opt. Lett.}, 42(19):3796--3799, 2017.

\bibitem{sanchez20167}
D~Sanchez, M~Hemmer, M~Baudisch, SL~Cousin, K~Zawilski, P~Schunemann, O~Chalus,
  C~Simon-Boisson, and J~Biegert.
\newblock 7 $\mu$m, ultrafast, sub-millijoule-level mid-infrared optical
  parametric chirped pulse amplifier pumped at 2 $\mu$m.
\newblock {\em Optica}, 3(2):147--150, 2016.

\bibitem{von2015picosecond}
Lorenz von Grafenstein, Martin Bock, Dennis Ueberschaer, Uwe Griebner, and
  Thomas Elsaesser.
\newblock Picosecond 34 mj pulses at khz repetition rates from a ho: Ylf
  amplifier at 2 $\mu$m wavelength.
\newblock {\em Opt. Express}, 23(26):33142--33149, 2015.

\bibitem{von2016ho}
L~von Grafenstein, M~Bock, D~Ueberschaer, U~Griebner, and T~Elsaesser.
\newblock Ho: Ylf chirped pulse amplification at kilohertz repetition
  rates--4.3 ps pulses at 2 $\mu$m with gw peak power.
\newblock {\em Opt. Lett.}, 41(20):4668--4671, 2016.

\bibitem{malevich2016broadband}
Pavel Malevich, Tsuneto Kanai, Heinar Hoogland, Ronald Holzwarth, Andrius
  Baltu{\v{s}}ka, and Audrius Pug{\v{z}}lys.
\newblock Broadband mid-infrared pulses from potassium titanyl arsenate/zinc
  germanium phosphate optical parametric amplifier pumped by tm,
  ho-fiber-seeded ho: Yag chirped-pulse amplifier.
\newblock {\em Opt. Lett.}, 41(5):930--933, 2016.

\bibitem{haberberger2010fifteen}
D~Haberberger, S~Tochitsky, and C~Joshi.
\newblock Fifteen terawatt picosecond co 2 laser system.
\newblock {\em Opt. Express}, 18(17):17865--17875, 2010.

\bibitem{polyanskiy2011picosecond}
Mikhail~N Polyanskiy, Igor~V Pogorelsky, and Vitaly Yakimenko.
\newblock Picosecond pulse amplification in isotopic co 2 active medium.
\newblock {\em Opt. Express}, 19(8):7717--7725, 2011.

\bibitem{pupeza2015high}
Ioachim Pupeza, D~S{\'a}nchez, Jinwei Zhang, Nicolai Lilienfein, Marcus Seidel,
  Nicholas Karpowicz, T~Paasch-Colberg, Irina Znakovskaya, M~Pescher,
  W~Schweinberger, et~al.
\newblock High-power sub-two-cycle mid-infrared pulses at 100 mhz repetition
  rate.
\newblock {\em Nat. Photonics}, 9(11):721--724, 2015.

\bibitem{krogen2017generation}
Peter Krogen, Haim Suchowski, Houkun Liang, Noah Flemens, Kyung-Han Hong,
  Franz~X K{\"a}rtner, and Jeffrey Moses.
\newblock Generation and multi-octave shaping of mid-infrared intense
  single-cycle pulses.
\newblock {\em Nat. Photonics}, 11(4):222--226, 2017.

\bibitem{nomura2012phase}
Yutaka Nomura, Hideto Shirai, Kenta Ishii, Noriaki Tsurumachi, Alexander~A
  Voronin, Aleksei~M Zheltikov, and Takao Fuji.
\newblock Phase-stable sub-cycle mid-infrared conical emission from
  filamentation in gases.
\newblock {\em Opt. Express}, 20(22):24741--24747, 2012.

\bibitem{fuji2015generation}
Takao Fuji, Yutaka Nomura, and Hideto Shirai.
\newblock Generation and characterization of phase-stable sub-single-cycle
  pulses at 3000 cm $^{-1}$.
\newblock {\em IEEE J. Sel. Top. Quantum Electron.}, 21(5):1--12, 2015.

\bibitem{pigeon2016measurements}
JJ~Pigeon, S~Ya Tochitsky, EC~Welch, and C~Joshi.
\newblock Measurements of the nonlinear refractive index of air, n 2, and o 2
  at 10 $\mu$m using four-wave mixing.
\newblock {\em Opt. Lett.}, 41(17):3924--3927, 2016.

\bibitem{junginger2010single}
Friederike Junginger, Alexander Sell, Olaf Schubert, Bernhard Mayer, Daniele
  Brida, M~Marangoni, Giulio Cerullo, Alfred Leitenstorfer, and Rupert Huber.
\newblock Single-cycle multiterahertz transients with peak fields above 10
  mv/cm.
\newblock {\em Opt. Lett.}, 35(15):2645--2647, 2010.

\bibitem{silva2012multi}
F~Silva, Dane~R Austin, A~Thai, M~Baudisch, M~Hemmer, D~Faccio, A~Couairon, and
  J~Biegert.
\newblock Multi-octave supercontinuum generation from mid-infrared
  filamentation in a bulk crystal.
\newblock {\em Nat. Commun.}, 3:807, 2012.

\bibitem{pigeon2014supercontinuum}
JJ~Pigeon, S~Ya Tochitsky, C~Gong, and C~Joshi.
\newblock Supercontinuum generation from 2 to 20 $\mu$m in gaas pumped by
  picosecond co2 laser pulses.
\newblock {\em Opt. Lett.}, 39(11):3246--3249, 2014.

\bibitem{mitrofanov2016subterawatt}
AV~Mitrofanov, AA~Voronin, DA~Sidorov-Biryukov, SI~Mitryukovsky, AB~Fedotov,
  EE~Serebryannikov, DV~Meshchankin, V~Shumakova, S~Ali{\v{s}}auskas,
  A~Pug{\v{z}}lys, et~al.
\newblock Subterawatt few-cycle mid-infrared pulses from a single filament.
\newblock {\em Optica}, 3(3):299--302, 2016.

\bibitem{shumakova2016multi}
V~Shumakova, P~Malevich, S~Ali{\v{s}}auskas, A~Voronin, AM~Zheltikov, D~Faccio,
  D~Kartashov, A~Baltu{\v{s}}ka, and A~Pug{\v{z}}lys.
\newblock Multi-millijoule few-cycle mid-infrared pulses through nonlinear
  self-compression in bulk.
\newblock {\em Nat. Commun.}, 7, 2016.

\bibitem{liang2017high}
Houkun Liang, Peter Krogen, Zhou Wang, Hyunwook Park, Tobias Kroh, Kevin
  Zawilski, Peter Schunemann, Jeffrey Moses, Louis~F DiMauro, Franz~X
  K{\"a}rtner, et~al.
\newblock High-energy mid-infrared sub-cycle pulse synthesis from a parametric
  amplifier.
\newblock {\em Nat. Commun.}, 8(1):141, 2017.

\bibitem{gordon2003asymmetric}
DF~Gordon, B~Hafizi, RF~Hubbard, JR~Penano, P~Sprangle, and A~Ting.
\newblock Asymmetric self-phase modulation and compression of short laser
  pulses in plasma channels.
\newblock {\em Phys. Rev. Lett.}, 90(21):215001, 2003.

\bibitem{tsung2002generation}
FS~Tsung, C~Ren, LO~Silva, WB~Mori, and T~Katsouleas.
\newblock Generation of ultra-intense single-cycle laser pulses by using photon
  deceleration.
\newblock {\em Proceedings of the National Academy of Sciences}, 99(1):29--32,
  2002.

\bibitem{sprangle1990nonlinear0}
P~Sprangle, E~Esarey, and A~Ting.
\newblock Nonlinear theory of intense laser-plasma interactions.
\newblock {\em Phys. Rev. Lett.}, 64(17):2011, 1990.

\bibitem{sprangle1990nonlinear}
P~Sprangle, E~Esarey, and A~Ting.
\newblock Nonlinear interaction of intense laser pulses in plasmas.
\newblock {\em Phys. Rev. A}, 41(8):4463, 1990.

\bibitem{wilks1989photon}
SC~Wilks, JM~Dawson, WB~Mori, T~Katsouleas, and ME~Jones.
\newblock Photon accelerator.
\newblock {\em Phys. Rev. Lett.}, 62(22):2600, 1989.

\bibitem{esarey1990frequency}
E~Esarey, A~Ting, and P~Sprangle.
\newblock Frequency shifts induced in laser pulses by plasma waves.
\newblock {\em Phys. Rev. A}, 42(6):3526, 1990.

\bibitem{mori1997physics}
WB~Mori.
\newblock The physics of the nonlinear optics of plasmas at relativistic
  intensities for short-pulse lasers.
\newblock {\em IEEE J. Quantum Electron.}, 33(11):1942--1953, 1997.

\bibitem{pai2010generation}
C-H Pai, Y-Y Chang, L-C Ha, Z-H Xie, M-W Lin, J-M Lin, Y-M Chen, G~Tsaur, H-H
  Chu, S-H Chen, et~al.
\newblock Generation of intense ultrashort midinfrared pulses by laser-plasma
  interaction in the bubble regime.
\newblock {\em Phys. Rev. A}, 82(6):063804, 2010.

\bibitem{guenot2017relativistic}
D~Gu{\'e}not, D~Gustas, A~Vernier, B~Beaurepaire, F~B{\"o}hle, M~Bocoum,
  M~Lozano, A~Jullien, R~Lopez-Martens, Agustin Lifschitz, et~al.
\newblock Relativistic electron beams driven by khz single-cycle light pulses.
\newblock {\em Nat. Photonics}, 11(5):293--296, 2017.

\bibitem{fonseca2002osiris}
Ricardo~A Fonseca, Lu{\'\i}s~O Silva, Frank~S Tsung, Viktor~K Decyk, Wei Lu,
  Chuang Ren, Warren~B Mori, S~Deng, S~Lee, T~Katsouleas, et~al.
\newblock Osiris: a three-dimensional, fully relativistic particle in cell code
  for modeling plasma based accelerators.
\newblock In {\em International Conference on Computational Science}, pages
  342--351. Springer, 2002.

\bibitem{fonseca2008one}
RA~Fonseca, SF~Martins, LO~Silva, JW~Tonge, FS~Tsung, and WB~Mori.
\newblock One-to-one direct modeling of experiments and astrophysical
  scenarios: pushing the envelope on kinetic plasma simulations.
\newblock {\em Plasma Phys. Control. Fusion}, 50(12):124034, 2008.

\bibitem{xu2017high}
XL~Xu, F~Li, W~An, TN~Dalichaouch, P~Yu, W~Lu, C~Joshi, and WB~Mori.
\newblock High quality electron bunch generation using a longitudinal
  density-tailored plasma-based accelerator in the three-dimensional blowout
  regime.
\newblock {\em Phys. Rev. Accel. Beams}, 20(11):111303, 2017.

\bibitem{guillaume2015electron}
Emilien Guillaume, Andreas D{\"o}pp, C{\'e}dric Thaury, K~Ta Phuoc,
  A~Lifschitz, G~Grittani, J-P Goddet, A~Tafzi, Shao-Wei Chou, L{\'a}szl{\'o}
  Veisz, et~al.
\newblock Electron rephasing in a laser-wakefield accelerator.
\newblock {\em Phys. Rev. Lett.}, 115(15):155002, 2015.

\bibitem{ralph2009self}
JE~Ralph, KA~Marsh, AE~Pak, W~Lu, CE~Clayton, F~Fang, WB~Mori, and C~Joshi.
\newblock Self-guiding of ultrashort, relativistically intense laser pulses
  through underdense plasmas in the blowout regime.
\newblock {\em Phys. Rev. Lett.}, 102(17):175003, 2009.

\bibitem{lifschitz2009particle}
AF~Lifschitz, X~Davoine, E~Lefebvre, J{\'e}r{\^o}me Faure, Cl{\'e}ment
  Rechatin, and Victor Malka.
\newblock Particle-in-cell modelling of laser--plasma interaction using fourier
  decomposition.
\newblock {\em J. Comput. Phys.}, 228(5):1803--1814, 2009.

\bibitem{davidson2015implementation}
Adam Davidson, Adam Tableman, Weiming An, Frank~S Tsung, Wei Lu, Jorge Vieira,
  Ricardo~A Fonseca, Lu{\'\i}s~O Silva, and Warren~B Mori.
\newblock Implementation of a hybrid particle code with a pic description in
  r--z and a gridless description in $\phi$ into osiris.
\newblock {\em J. Comput. Phys.}, 281:1063--1077, 2015.

\end{thebibliography}

\section*{Acknowledgements}
This work was supported by the NSFC Grants No. 11425521, No. 11535006, No. 11475101 and No. 11775125, Thousand Young Talents Program, the Air Force Office of Scientific Research (AFOSR) under award number FA9550-16-1-0139 DEF, the Office of Naval Research (ONR) MURI (4-442521-JC-22891) and by the U.S. Department of Energy grant DE-SC001006, and the Ministry of Science and Technology of Taiwan under Grants No. MOST-105-2112-M-001-005-M3. The simulations were performed on Sunway TaihuLight.

\section*{Author contributions}
Z.N., C.H.P. and W.L. proposed the concept. Z.N. developed the theoretical model, carried out the simulations, and wrote the paper. C.H.P., J.F.H., C.J.Z., Y.W.P., Y.W., F.L., W.L., and C.J. contributed to refining the details of the paper. All authors reviewed the manuscript.

\end{document}